\documentclass[prb,twocolumn,showpacs,preprintnumbers,amsmath,amssymb,superscriptaddress,floatfix]{revtex4-2}

\usepackage{amsmath}    % need for subequations
\usepackage{amssymb,environ}
\usepackage{graphicx}   % need for figures
\usepackage{verbatim}   % useful for program listings
\usepackage{color}      % use if color is used in text
\usepackage{hyperref}   % use for hypertext links, including those to external documents and URLs
\usepackage{breakurl}
\usepackage[normalem]{ulem}
\usepackage{natbib}
\usepackage{enumitem}
\usepackage{multirow}

\usepackage{epsfig}
\usepackage{textcomp}
\usepackage{wasysym}
\usepackage{array}
\usepackage{color}

\begin{document}

\newcommand{\ev}[0]{\mathbf{e}}
\newcommand{\cv}[0]{\mathbf{c}}
\newcommand{\fv}[0]{\mathbf{f}}
\newcommand{\Rv}[0]{\mathbf{R}}
\newcommand{\Tr}[0]{\mathrm{Tr}}
\newcommand{\ud}[0]{\uparrow\downarrow}
\newcommand{\Uv}[0]{\mathbf{U}}
\newcommand{\Iv}[0]{\mathbf{I}}
\newcommand{\Hv}[0]{\mathbf{H}}

\setlength{\jot}{2mm}

\newcommand{\jav}[1]{#1}

\title{Non-equilibrium dynamics of superconductivity in the Hatsugai-Kohmoto model}

\author{\'Ad\'am B\'acsi}
\email{bacsi.adam@sze.hu}
\affiliation{Department of Mathematics and Computational Sciences, Sz\'echenyi Istv\'an University, 9026 Gy\H or, Hungary}
%\affiliation{Jo\v zef Stefan Institute, Jamova 39, Ljubljana SI-1000, Slovenia}

\author{Bal\'azs D\'ora}
\email{dora.balazs@ttk.bme.hu}
%\affiliation{MTA-BME Lend\"ulet Topology and Correlation Research Group,Budapest University of Technology and Economics, 1521 Budapest, Hungary}
\affiliation{Department of Theoretical Physics, Budapest University of Technology and Economics, Budapest, Hungary}

\date{\today}

\begin{abstract}
We study the non-equilibrium dynamics of the superconducting order parameter in the Hatsugai-Kohmoto (HK) model. In the absence of superconductivity, its ground state is a non-Fermi liquid, whose
properties are controlled by the HK interaction. Our protocol involves quantum quenching the HK interaction but leaving the 
interaction responsible for superconductivity unchanged.
We map out the non-equilibrium dynamical phase diagram of the interacting model which contains three phases where, at long times, 
the order parameter amplitude vanishes, approaches a constant value or persistently oscillates.
We also investigate the Loschmidt echo in searching for dynamical quantum phase transition, and 
find \jav{that its non-analytic temporal behaviour is close but does not match exactly the vanishing of the order parameter}.
The momentum space entanglement entropy  between positive and negative momentum modes, relevant for \jav{Cooper} pairing, is calculated.
Counterintuitively, this momentum space entanglement does not change significantly during the quench dynamics and its value remains reasonably large even for  vanishing superconducting
order parameter. Nevertheless, its derivative with respect to the  HK interaction signals the dynamical phase transition associated to the late time vanishing of superconductivity.

\end{abstract}

\maketitle
\section{Introduction}
The investigation of quantum quenches and non-equilibrium dynamics in general represents an important  tool in several distinct fields of physics, including condensed matter, cold atoms, high energy physics, mesoscopic systems etc.\cite{dziarmagareview,polkovnikovrmp,annurevmitra}.
During the non-equilibrium time evolution, one can on one hand address important and fundamental
questions related to equilibration and thermalization, but on the other hand, engineering novel steady states of matter without an equilibrium counterpart is
possible, such as e.g. Floquet topological systems without an external drive\cite{foster}.

This program has been carried out for interacting superconductors upon an abrupt temporal change of the interaction strength, responsible for superconductivity.
In general, three different phases are distinguished based on the behaviour of the order parameter in the long time limit: 
the superconducting order parameter vanishes, approaches a constant value or persistently oscillates \cite{barankov2,young}.
The dynamical phase transition (DPT) between these phases is continuous and occurs as the initial or final interaction is varied\cite{Marino2022}.
Parallel to these, another notion of dynamical quantum phase transition (DQPT) appeared, which is related to the non-analytic behaviour in time of the Loschmidt echo, namely the overlap
of the initial and time evolved wavefunctions\cite{Heyl_2018,kehrein2013,pollmannpre}. For quenched BCS superconductors, the temporal non-analyticity  
in the Loschmidt echo does not distinguish between 
distinct dynamical phases\cite{yuzbashyan2021}.

The common feature in these non-equilibrium BCS superconductors is the Fermi-liquid nature of the normal, non-superconducting ground state. However, recently equilibrium 
superconductivity\cite{Phillips2020,zaanen,Li2022,Zhu2021,zhuprb2021,zhao2022,Sun2024} arising on top
of the competition between non-Fermi liquid behaviour and Mottness has been studied in the Hatsugai-Kohmoto (HK) model\cite{hatsugai1992,lidsky}.
Therein, the special structure of the momentum space interaction renders the model exactly solvable and induces non-Fermi liquid behaviour for a wide range of model parameters.

Here we extend previous studies on non-equilibrium interacting BCS superconductivity based on the HK model. In contrast to previous approaches, we study 
the effect of quantum  quenches of the HK interaction on the superconducting order parameter in the initially superconducting HK. We also investigate the conditions 
for emerging dynamical phase transitions (DPT).
We find that for a HK quench, the long time phase diagram has an entirely different structure compared to the conventional case\cite{barankov,barankov2,gurarie,yuzbashyan2006} with Fermi-liquid normal state, though
the ensuing phases display the same three long time behaviours, namely the amplitude of the superconducting order parameter either vanishes or approaches a constant value or persistently oscillates.
In particular, we find that by significantly increasing or decreasing the HK interaction during the quench completely suppresses and kills 
superconductivity
in the long time limit, while small changes in the HK interaction result in persistent oscillation of the
superconducting order parameter.
In between these two extreme regions, the order parameter approaches a constant value\cite{volkov} through damped oscillations.  
By investigating the Loschmidt echo, we find no direct correspondence between the non-analyticity of the echo and the order parameter phase diagram. 

We also investigated the reduced density matrix of positive momentum modes and the ensuing entanglement entropy between positive and negative momentum modes.
We find 
%no strong changes in its overall behaviour and 
that
it remains sizeable even for vanishing late time superconducting order parameter. This indicates that although superconductivity
is absent from certain final states, superconducting quantum correlations are still relevant.

\jav{The paper is structured as follows. After introducing the superconducting HK model in Sec. \ref{sec:HKmod} and describing the interaction quench in Sec. \ref{sec:quench}, we present the results on the superconducting order paremeter in Sec. \ref{sec:ordp}. In Sec. \ref{sec:LE}, the non-analytic features of the LE are studied, while in Sec. \ref{sec:ent}, the time evolution of the momentum space entanglement is presented. }

\section{Superconducting Hatsugai-Kohmoto model}
\label{sec:HKmod}
The HK model \cite{hatsugai1992,Phillips2020} is defined by the Hamiltonian
\begin{gather}
H_0=\sum_k\Big( \varepsilon(k) (n_{k\uparrow} + n_{k\downarrow}) + U n_{k\uparrow} n_{k\downarrow}\Big)
\end{gather}
where $n_{k\sigma} = c_{k\sigma}^+ c_{k\sigma}$ is the occupation number operator of the electrons with momentum $k$ and spin $\sigma$ and $\varepsilon(k)$ is the non-interacting band structure. 
The dimension of the system does not play any qualitative role in the forthcoming calculations and, hence, the dimension of the wavevector $k$ is also not specified. For each momentum $k$, 
the Hamiltonian is diagonalizable in the four-dimensional many-body Hilbert space spanned by $|0\rangle$, $|\uparrow\rangle$, $|\downarrow\rangle$ and $|\ud\rangle$ denoting the states 
with zero particle, one spin up particle, one spin down particle and two particles, respectively. The eigenvalues are 
$0$ for the empty state, $\varepsilon(k)$ for singly occupied state with either a spin up or a spin down particle, and 
finally $2\varepsilon(k) + U$ for the doubly occupied case. The band structure is best represented 
by a lower ($\varepsilon(k)$) and an upper Hubbard band ($\varepsilon(k) + U$) \cite{Phillips2020}. 
\jav{The ground state is a non-Fermi liquid since the elementary excitations are not fermionic. An elementary excitation in the lower Hubbard band,
$\varepsilon(k)$ band is created by $c_{k\sigma}^+(1-n_{k \bar\sigma})$,
while excitations in the upper Hubbard band are of the form $c_{k\sigma}^+ n_{k \bar\sigma}$. These do not satisfy fermionic algebra but rather resemble
holon and doublon composite excitations.}
The two Hubbard bands overlap if $U<W$ and are completely separated for $U\geq W$ where $W$ is the 
bandwidth of the dispersion $\varepsilon(k)$, i.e., $\varepsilon\in \left(-W/2, W/2\right)$. At zero temperature, the bands are filled up to the chemical potential $\mu$. The half filling is ensured by the relation 
\begin{gather}
\mu = \frac U2
\label{chempot}
\end{gather}
 if the bands obey the symmetry of $\rho(\varepsilon) = \rho(-\varepsilon)$ where $\rho$ is the density of states of the energy levels $\varepsilon$. 
For half filling, the ground state of the system is a non-Fermi liquid for $U<W$ and the amount of non-Fermi liquidness is controlled by $U$, 
while it becomes a Mott insulator for $U>W$\cite{Phillips2020}.

To study superconductivity in the HK model, we consider an attractive interaction \cite{Phillips2020,Li2022,Zhu2021} of the form
\begin{gather}
H= H_0 - \frac{g}{N}\sum_{kk'}c^+_{k\uparrow} c_{-k\downarrow}^+ c_{-k'\downarrow} c_{k'\uparrow}
\end{gather}
where $g$ is the pairing strength and $N$ is the number of momentum modes.
Within mean-field approximation, we assume that the fluctuations around the expectation value $\Delta = \frac{g}{N}\sum_{k}\langle c_{-k\downarrow}c_{k\uparrow}\rangle$ are negligible, leading to
\begin{gather}
H_{MF} = H_0 + \frac{N|\Delta|^2}{g} - \sum_{k} \left( \Delta c^+_{k\uparrow} c_{-k\downarrow}^+ + h.c.\right)\,.
\label{hmf}
\end{gather}
The pairing interaction connects the sectors corresponding to $k$ and $-k$. The overall many-body Hilbert  space for a given $(k,-k)$ becomes 16-dimensional \cite{Zhu2021}. 
In this Hilbert space, all possible occupation configurations of the states $k\uparrow$, $k\downarrow$, $-k\uparrow$ and $-k\downarrow$ are taken into 
account. By considering the ensuing pseudo-spins \cite{anderson1958,barankov2} of these four states, the direct product of the four spin-$\frac{1}{2}$ space results 
in the direct sum of a quintet subspace (spin-$2$), 3 triplet subspaces (spin-$1$) and 2 singlet subspaces (spin-$0$) as
\begin{gather}
\mathcal{H}_{\frac{1}{2}}\otimes \mathcal{H}_{\frac{1}{2}} \otimes \mathcal{H}_{\frac{1}{2}} \otimes \mathcal{H}_{\frac{1}{2}} = 2\mathcal{H}_0 \oplus 3\mathcal{H}_1 \oplus \mathcal{H}_2,
\end{gather}
where $\mathcal{H}_{S}$ denotes the local Hilbert space of a spin-$S$.
The pseudo-spin operators are defined as
\begin{subequations}
\begin{gather}
\hat{S}_{-,k} = b_k = \hat{S}_{+,k}^+,  \\
\hat{S}_{z,k} = \frac{n_k}{2}-1,  \\
\hat{S}^2_k = \left(\frac{n_k}{2}-1\right)^2 + \frac{b_k^+ b_k + b_k b_k^+}{2},
\end{gather}
\label{pseudospin}
\end{subequations}
where $n_k = \sum_{\sigma} \left(n_{k\sigma}+n_{-k\sigma}\right)$ is the operator counting the total number of particles in the modes $k$ and $-k$ and
\begin{gather}
b_k = c_{-k\downarrow} c_{k\uparrow} + c_{k\downarrow} c_{-k\uparrow}
\end{gather}
is the operator annihilating a Cooper-pair. 
By analyzing the Hamiltonian in a given $k>0$ mode, the lowest energy is found in the three-dimensional pseudo-spin triplet subspace spanned by
\begin{subequations}
\begin{gather}
|0_k\rangle = |0,0,0,0\rangle,  \\ 
|S_k\rangle = \frac{1}{\sqrt 2}b_k^+ |0_k\rangle = \nonumber\\
 = \frac{1}{\sqrt{2}}\left(|\uparrow_k,\downarrow_{-k},0,0\rangle + |0,0,\uparrow_{-k},\downarrow_k\rangle\right),\\
|D_k\rangle =|\uparrow_k,\downarrow_{-k},\uparrow_{-k},\downarrow_k \rangle =\frac{1}{\sqrt 2} b_k^+ |S_k\rangle
\end{gather}
\label{eq:singstates}
\end{subequations}
which, from the point of view of the physical spin, are singlet states.  Eqs. \eqref{eq:singstates} represents the many-body bases compared to the conventional BCS case, wherein the respective states contain 0, 2 and 4 particles.
From now on, we will consider only half of the first Brillouin zone instead of all $k$ modes and use the notation $k>0$ for the half Brillouin zone independently from the actual dimension of the system. The number of $k>0$ modes is $N/2$.
On the subspace of the states in Eq. \eqref{eq:singstates}, the grand canonical Hamiltonian $H_{MF}-\mu \hat{N}$ is represented by
\begin{gather}
\Hv_k = \left[ \begin{array}{ccc} 0 & -\sqrt{2}\Delta^* & 0 \\ -\sqrt{2} \Delta & 2\varepsilon(k) - 2\mu & -\sqrt{2}\Delta^* \\ 0 & -\sqrt{2}\Delta & 4\varepsilon(k) - 4\mu + 2U \end{array}\right]
\label{eq:hams}
\end{gather}
where we assumed that $\varepsilon(k) = \varepsilon(-k)$. The effective Hamiltonian in Eq. \eqref{eq:hams} is in accordance with Ref. \cite{Zhu2021}. 
Eq. \eqref{eq:hams} can in principle be diagonalized, but its eigenvalues are only simple expressions in the uncorrelated, $U=0$ case as 
$2(\varepsilon(k) - \mu)+2s\sqrt{(\varepsilon(k) - \mu)^2+|\Delta|^2}$
with $s=-1$, 0 and 1.
We finally mention that that the  non-Fermi liquid non-superconducting ground state of the HK model has huge degeneracy\cite{Phillips2020}, 
but the extra pairing interaction $g$ lifts this  ground state degeneracy and results in a non-degenerate superconducting ground state.

%Based on the results of Ref. \cite{Zhu2021}, one can numerically obtain the ground state phase diagram as a function of $U$ and $g$. Fig. \ref{fig:phdiag} shows the value of the gap in the ground state of the system. For the phase diagram, constant density of states is considered between $-W/2$ and $W/2$.

\begin{figure}[h]
\centering
\includegraphics[width=7cm]{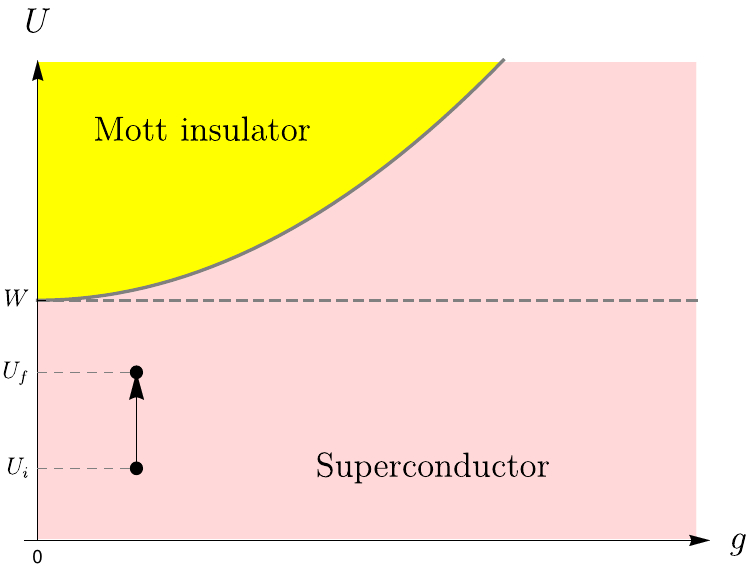}
\caption{Schematic of the mean-field phase diagram at $T=0$ and at half filling. In the absence of superconductivity ($g=0$), the system forms a non-Fermi liquid for $U<W$.
The arrow indicates the sudden quench of the interaction strength studied in Sec. \ref{sec:quench}.}
\label{fig:phdiag}
\end{figure}

Within mean-field approximation, the ground state phase diagram at half-filling is sketched in Fig. \ref{fig:phdiag}. When the two Hubbard bands overlap, $U<W$, 
the superconducting phase is favoured for any finite pairing interaction $g$. 
In the following section, we study the dynamics following a sudden quench of the HK interaction $U$ inside this regime by keeping $g$ fixed, unchanged. 
This means that we change the non-Fermi liquid properties of the non-interacting system by quenching $U$ and ask, how this affects the already existing superconducting properties.
Non-equilibrium dynamics of the non-superconducting HK model was also investigated\cite{nesselrodt} for the  photoemission spectra. 

\section{Sudden quench of the interaction strength}
\label{sec:quench}
The system is initially prepared in the ground state corresponding to a pairing interaction $g$ and initial $U_i$ in the equilibrium non-Fermi liquid phase. 
As shown in the previous section, the ground state is a singlet state corresponding to the total spin zero. 
At $t=0$, the strength of $U$ is changed from $U_i$ to $U_f$ within the non-Fermi liquid region, namely $0<U_{i,f}<W$. 
The resulting Hamiltonian conserves the total and the pseudospin as well  and, hence, 
the dynamics will not be driven out of the pseudospin triplet subspace, spanned by Eq. \eqref{eq:singstates}.

The most general form of the singlet wavefunction can be written in the form of
\begin{gather}
|\Psi(t)\rangle = \prod_{k>0}\left(x_k(t) |0_k\rangle + y_k(t)|S_k\rangle + z_k(t)  |D_k\rangle\right)
\end{gather}
where the coefficients fulfill $|x_k(t)|^2 + |y_k(t)|^2 + |z_k(t)|^2 =1$ at any time instants. 
The conventional uncorrelated, $U=0$ BCS theory follows from $x_k(t)=u_k^2(t)$, $y_k(t)=\sqrt{2} u_k(t)v_k(t)$ and $z_k(t)=v_k^2(t)$ with $u_k(t)$ and $v_k(t)$ the conventional
Bogoliubov coefficients\cite{tinkham}, satisfying $|u_k(t)|^2+|v_k(t)|^2=1$.
The dynamics is governed by the Hamiltonian in Eq. \eqref{eq:hams}, resulting in the dynamical equations for the wavefunction coefficients as
\begin{gather}
%i\partial_t \left[\begin{array}{c} x_k(t) \\ y_k(t) \\ z_k(t) \end{array}\right] = \left[ \begin{array}{ccc} 0 & -\sqrt{2}\Delta(t)^* & 0 \\ -\sqrt{2} \Delta(t) & 2\varepsilon(k) - U_f & -\sqrt{2}\Delta^*(t) \\ 0 & -\sqrt{2}\Delta(t) & 4\varepsilon \end{array}\right]
%\left[\begin{array}{c} x_k(t) \\ y_k(t) \\ z_k(t) \end{array}\right]
%i\partial_t \phi(t) = \left[ \begin{array}{ccc} 0 & -\sqrt{2}\Delta(t)^* & 0 \\ -\sqrt{2} \Delta(t) & 2\varepsilon(k) - U_f & -\sqrt{2}\Delta^*(t) \\ 0 & -\sqrt{2}\Delta(t) & 4\varepsilon \end{array}\right]\phi(t)
i\partial_t \psi(t) =\Hv_k \psi(t)
\label{eq:diff}
\end{gather}
where $\psi(t)=[x_k(t), y_k(t), z_k(t)]^T$ and we used $U_f$ in $\Hv_k$ together with $\mu=U_f/2$, corresponding to half filling. 
The quench in the interaction strength also involves a quench in the chemical potential in order to not change the filling of the system.
Note that in Eq. \eqref{eq:diff}, the gap is time-dependent. Within the realm of self-consistent mean-field theory, the gap is calculated based on the instantaneous coefficients as
\begin{gather}
\Delta(t) = \frac{g}{N}\sum_{k>0} \sqrt{2}\left(x_k^*(t) y_k(t) + y_k(t)^* z_k(t)\right)
\label{eq:gap}
\end{gather}
which is an extension of the calculation presented in Refs. \cite{barankov,barankov2,peronaci} to the HK model.
The initial condition for Eq. \eqref{eq:diff} is determined by the lowest energy eigenvectors of Eq. \eqref{eq:hams} 
when the interaction strength is set to $U_i$ and $\mu = U_i/2$ from Eq. \eqref{chempot}. 

Since the Hamiltonian does not commute with the number of particles, it is also a question if the filling factor remains $1/2$. It can, however, be shown that the expectation value of the number of particles is expressed as
\begin{gather}
N_e(t) = \sum_{k>0} \left(2 |y_k|^2 + 4|z_k|^2\right)
\label{eq:ne}
\end{gather}
and its time derivative is zero under the dynamics governed by Eq. \eqref{eq:diff}, i.e., the half-filling is preserved.
%\begin{gather}
%\frac{\mathrm{d} N_e}{\mathrm{d}t} = \frac{2\sqrt{2}}{i}\sum_{k>0}\left[ \Delta (-y_k^* x_k + y_k z_k^*) + \Delta^* (y_k x_k^* - y_k^* z_k)\right] = 0
%\end{gather}

%Furthermore, the coefficients obey the normalization condition $|x_k|^2 + |y_k|^2 +|z_k|^2 =1$. By using this relation and extending the idea of Ref. \cite{barankov}, we introduce two complex variables instead of $x_k$, $y_k$ and $z_y$ as
%\begin{gather}
%w_k(t) = \frac{x_k(t)}{y_k(t)} \qquad v_k(t) = \frac{z_k(t)}{y_k(t)}
%\end{gather}
%for which the dynamical equations become non-linear as given by
%\begin{gather}
%i\hbar \dot{w}_k = -\sqrt{2} \Delta(t)^* + \sqrt{2}\Delta(t) w_k^2 - (2\varepsilon(k)-U)w_k - \sqrt{2}\Delta(t)^* w_kv_k \nonumber \\
%i\hbar \dot{v}_k = \sqrt{2}\Delta(t) + \sqrt{2}\Delta(t) v_kw_k + (2\varepsilon(k) + U) v_k  - \sqrt{2}\Delta(t)^* v_k^2
%\label{eq:dyneq}
%\end{gather}
%and the gap is calculated by
%\begin{gather}
%\Delta(t) = \frac{g}{N}\sum_{k>0}\sqrt{2}\frac{w_k(t)^* - v_k(t)}{1 +|w_k(t)|^2 + |v_k(t)|^2}\,.
%\label{eq:delta}
%\end{gather}

\section{Order parameter dynamics and phase diagram}
\label{sec:ordp}

The differential equations in Eq. \eqref{eq:diff} cannot be solved analytically. Here we have three coupled differential equations compared to the conventional case involving only two \cite{barankov2,yuzbashyan2021}. However, the 
main difficulty stems from the self-consistency relation from Eq. \eqref{eq:gap} 
which introduces a highly non-linear coupling between the dynamical variables. Moreover, even the eigenvalue problem in Eq. \eqref{eq:hams} does not admit a simple, transparent analytical solution for 
nonzero $U$.
The dynamical equations are solved numerically using the explicit Runge-Kutta method of order 8(5,3) 
for $x_k(t)$, $y_k(t)$ and $z_k(t)$.

\begin{figure}[h]
\centering
\includegraphics[width=8cm]{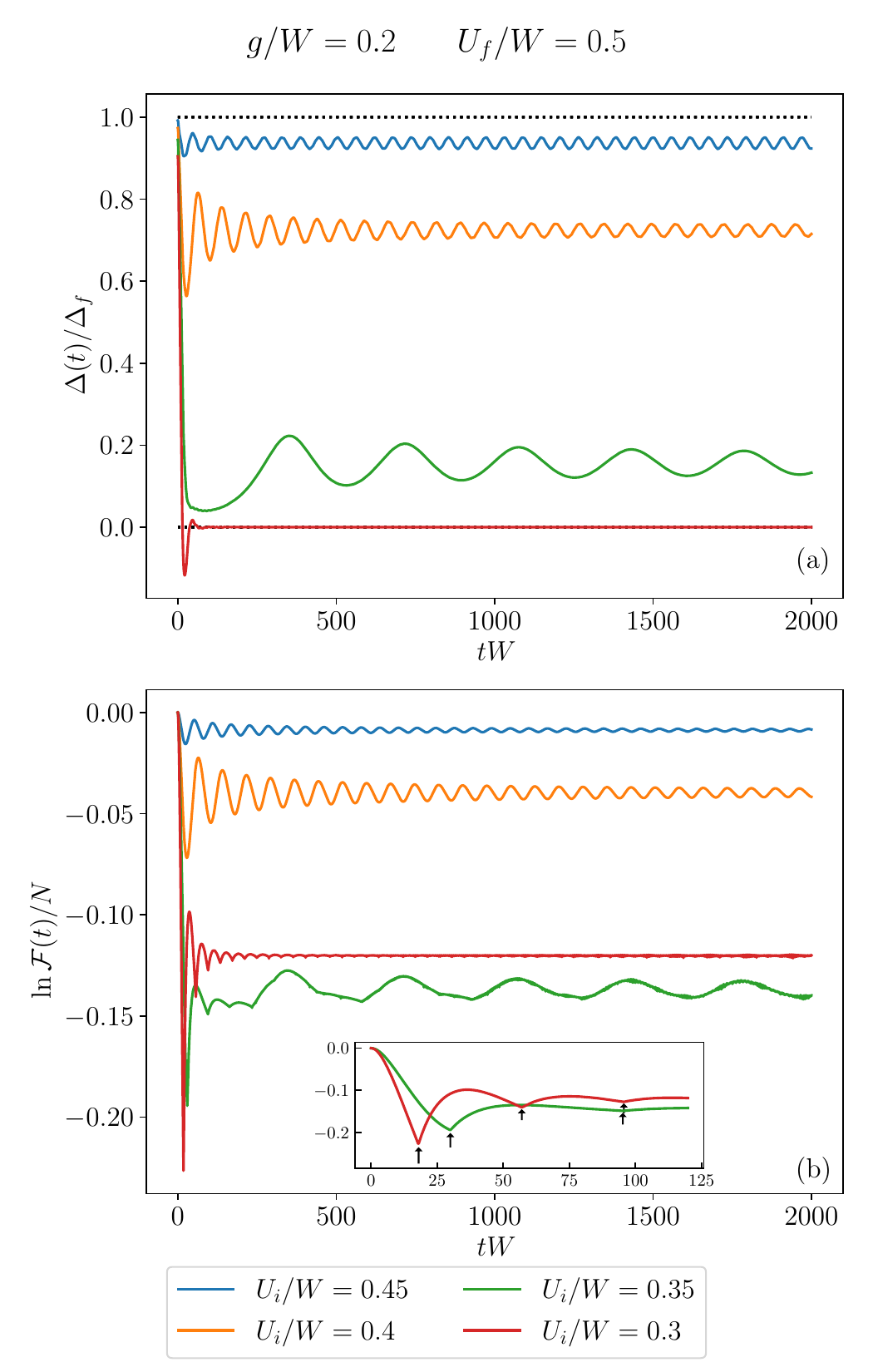}
\caption{Time evolution of \jav{a)} the order parameter, $\Delta(t)$, and \jav{b)} the Loschmidt echo, $\mathcal{F}(t)$ for various initial interaction strength, $U_i$. The green curve corresponding to $U_i/W=0.35$ denotes a non-zero late time superconducting order parameter, yet
non-analytic structures are already present in the Loschmidt echo. \jav{The inset zooms into the early time evolution of the Loschmidt echo and highlights the non-analytic features (cusps).}}
\label{fig:Uquench}
\end{figure}

\jav{From Eqs. \eqref{hmf} and \eqref{eq:hams}, we see that all $k$ dependencies occur only through $\varepsilon(k)$, therefore we can transform momentum sums into energy integrals using
the conventional trick as
$\sum_k f(\varepsilon(k))=N \int_{-W/2}^{W/2}d\varepsilon \rho(\varepsilon )f(\varepsilon)$ with $f(\varepsilon(k))$ corresponding to some physical quantity
as in e.g. Eq.  \eqref{eq:ne}, $\rho(\varepsilon)$ is the density of states what we approximate with $1/W$. Any other band structures can be taken into account by modifying the density of states.}
The total energy range $[-W/2;W/2]$ is discretized with the resolution $N_{res}=8000$, i.e. we consider $N_{res}$ equidistant 
energy values within the total energy range. \jav{By studying the effect of the energy resolution, we find no significant 
modification in the short time behavior. Relevant differences are found for long times, typically for $t \gtrsim \pi N_{res}/W$, i.e. for times much longer than the inverse of the level spacing.}
Other band structures with bounded density of states are not expected to yield qualitatively different results.
The order parameter is computed based on Eq. \eqref{eq:gap}, and the typical time evolutions exhibit three typical behaviors, see Fig. \ref{fig:Uquench} (a). 
When the difference between the initial and final interaction strength is small, the gap exhibits undamped, persistent oscillations around a finite value after a transient phase, see for example the quench corresponding to $U_i=0.45W$.

When the difference is intermediate, the gap converges to a finite value through damped oscillations, see for example the quench corresponding to $U_i=0.4W$ or $U_i=0.35W$. 
The third kind of behavior is observed when the difference between the initial and final interaction strengths is large. In this case, see for example the quench from $U_i=0.3W$, the order parameter converges to zero in an oscillating fashion with exponential time dependence
as $\Delta(t)\sim \exp(-\lambda t)$, where $\lambda$ is the Landau damping.

All of these temporal behaviors have also been witnessed after quenching the pairing interaction in regular BCS superconductors \cite{barankov2}. However, there are important differences  compared to the conventional BCS case. 
Undamped oscillations are observed if the final equilibrium gap is much larger than the initial one, while decaying order parameter is found in the opposite regime, i.e., when the final equilibrium gap is much smaller. 
Further difference is that in the BCS case, the time evolution is determined only by the ratio $\Delta_i/\Delta_f$, while in the HK model, more energy scales are present including e.g. $U_{i,f}$, which all influence the characteristics of the time evolution.

It is also interesting to note that the frequency of the oscillations in the order parameter is not solely determined by the final gap, $\Delta_f$, but it is influenced 
by the numerous energy scales including those of the initial state. In Fig. \ref{fig:Uquench} (a), for example, the final Hamiltonian is the same for all quenches and the 
only difference is in the initial state but the curves exhibit oscillations with different frequencies.

%For quenches when the gap decays to zero in an oscillating way, the order parameter vanishes at certain time instants. These time instants can be regarded as critical points of a dynamical phase transition (DPT) \cite{kehrein2013}. 

\begin{figure}[h]
\centering
\includegraphics[width=8cm]{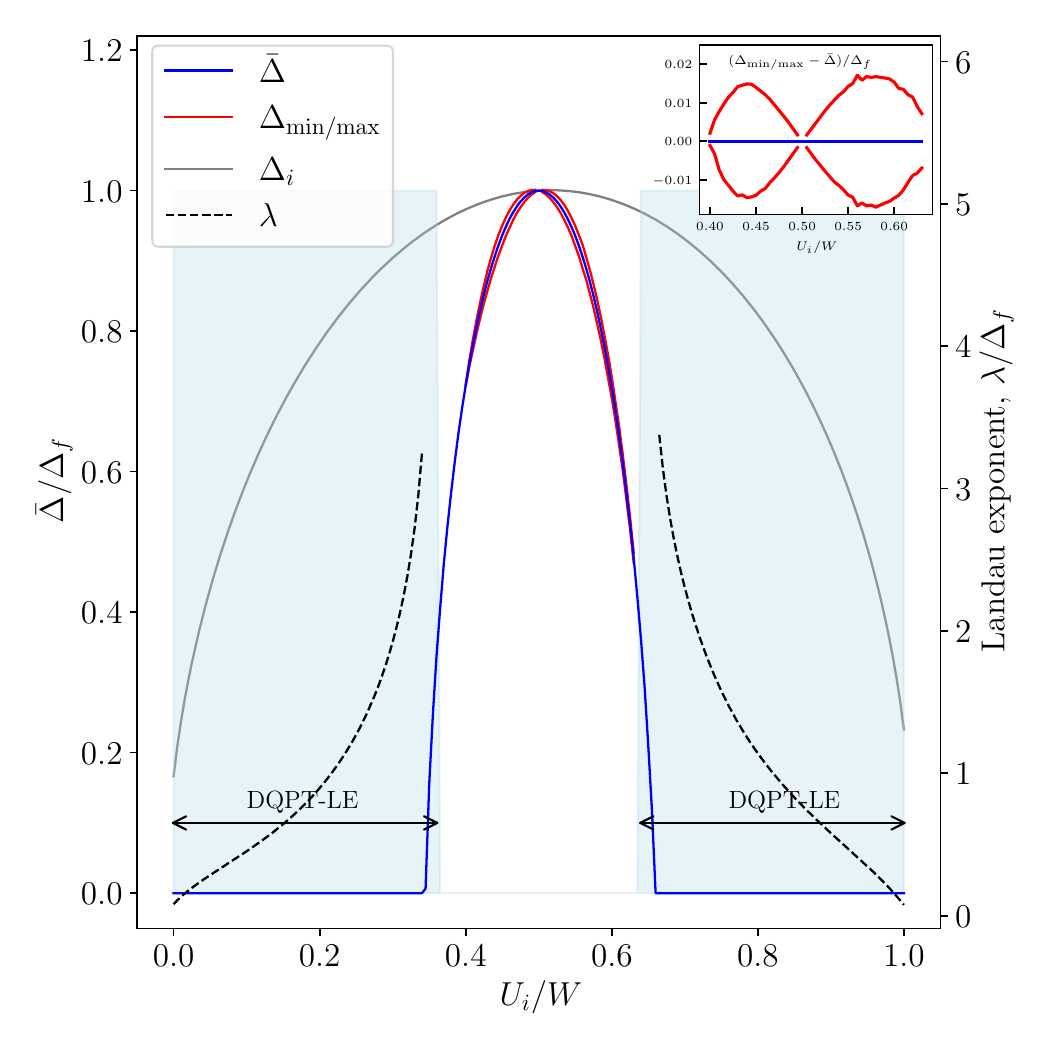}
\caption{The average order parameter (solid, blue line) for $U_f/W=0.5$ as a function of initial interaction strength $U_i$ measured on the left vertical axis, together with its maximal and minimal value in the case
of persistent oscillation. We also plot the equilibrium order parameter of the initial state (solid, gray line) for comparison, in accord with Refs. \cite{Phillips2020,Li2022,Zhu2021}.
The right vertical axis shows the Landau damping parameter (dashed, black lines) in the region when the order parameter vanishes in the long time limit. 
The \jav{shaded area} indicates the regime where the Loschmidt echo exhibits non-analytic features. It does not match the region of vanishing order parameter, the non-analyticities
already appear in the region when the long time limit of the superconducting order parameter approaches a non-zero constant.
For the plot, only the time instants after $t_\mathrm{cutoff} = 300 /W$ were taken into account.
The inset zooms into the region of persistent oscillation for small quenches.}
\label{fig:Uline}
\end{figure}

The cases, when the order parameter decays to zero, can be regarded as dynamical phase transition (DPT) \cite{zunkovic,Lerose}. 
To detect DPT numerically, we evaluate the average gap as
\begin{gather}
\bar{\Delta} = \frac{1}{N_{t_{\mathrm{cutoff}}}}\sum_{t_i>t_{\mathrm{cutoff}}}\Delta(t_i)
\end{gather}
where we introduced the temporal cut-off $t_{\mathrm{cutoff}}$ to avoid effects from the transient period. 
In the above expression, $N_{t_{\mathrm{cutoff}}}$ is the number of discrete time instants fulfilling $t_i>t_{\mathrm{cutoff}}$. In this way, $\bar{\Delta}$ describes the long term average of the order parameter. 
For the DPT case, i.e., when the gap decays to zero, the time average is approximately zero. The average order parameter as a function of $U_i$ is represented by the solid line in Fig. \ref{fig:Uline} for $g/W=0.2$ and $U_f/W=0.5$. 
At the DPT, $\bar{\Delta}=0$ and a sharp transition is detected at $U_i/W \approx 0.34$ and $U_i/W \approx 0.66$.

We further analyze the characteristics of the oscillations when the gap remains finite, i.e., in the absence of DPT. By extracting the local maxima and minima of the oscillations, we observe a power law decay with exponent close to -1/2
 of the amplitudes, in accord with the result of Ref. \onlinecite{volkov}. 
By fitting $\Delta_{\mathrm{min/max}} + A_{\mathrm{min/max}} t^{-1/2}$ separately to the local minima and maxima, 
we compute the long term $\Delta_{\mathrm{min/max}}$ values of the gap which are also plot in Fig. \ref{fig:Uline}. 
For comparison, we also show the HK interaction dependence of the initial superconducting order parameter.
It can be observed that $\Delta_{\mathrm{min/max}}$ coincide with $\Delta_f$ if $U_i = U_f$, i.e. no oscillations are present. 
As $U_i$ moves away from $U_f$, we first find persistent oscillations. This behavior then transforms smoothly to damped oscillations.
For large quenches, we find an exponential damping of the order parameter as $\Delta(t)\sim \exp(-\lambda t)$ with $\lambda$ the Landau damping. Its value is extracted from the time evolution
and is also plotted in Fig. \ref{fig:Uline}. 
While this phase diagram is not universal, namely the specific values depend on the chosen initial and final values of the interactions, we expect qualitatively similar features for other values  $g$ and $U_f$, see Appendix.

%As an alternative approach, we have also analyzed the Fourier transform of the timeseries $d_i$. Fig. \ref{fig:Umap} b) shows the frequency with the highest amplitude in the Fourier transform $\omega_r$. In the central region, where the average order parameter was found to be non-zero, the frequency with the highest amplitude is zero indicating that the constant term is the most dominant in the Fourier transform. When the difference between the initial and final interaction strengths is larger and the order parameter converges to zero, the frequency with the highest amplitude is a finite value. This approach shows the same region for DPT.

We note that in Ref. \cite{barankov2} the dynamical equations were written for the pseudospin components instead of $x_k$, $y_k$ and $z_k$. We choose to use these 
coefficients because they also enable to calculate other quantities such as the Loschmidt echo.

\section{Loschmidt echo}
\label{sec:LE}

Another definition of the dynamical quantum phase transition (DQPT) involves the time dependent non-analytic feature in the Loschmidt echo \cite{Heyl_2018, yuzbashyan2021,andraschko,karrasch}. To this end, we 
study the overlap between the time-evolved and the initial state, $\mathcal{F}(t) = |\langle \Psi_0 | \Psi(t) \rangle |^2$, where the time evolution involves the grand canonical Hamiltonian from Eq. \eqref{eq:hams}. 
By using the coefficients $x_k$, $y_k$ and $z_k$ from the wavefunction, the Loschmidt echo expressed as
\begin{gather}
\mathcal{F}(t) = \prod_{k>0}\left|x_k(0)^*x_k(t) + y_k(0)^*y_k(t) + z_k(0)^*z_k(t)\right|^2
\end{gather}
which can be computed directly from the numerical solution of the differential equations in Eq. \eqref{eq:diff}. 
For a conventional BCS non-equilibrium dynamics\cite{yuzbashyan2021}, this can be expressed in terms of the pseudospin expectation values from Eq. \eqref{pseudospin}, which does not work in our case when the non-superconducting ground state realizes a non-Fermi liquid.
In our case, the chemical potential is included in the Hamiltonian in Eq. \eqref{eq:hams} describing the dynamics of the system, and the ensuing time dependent overlap is the grand canonical Loschmidt 
echo\cite{yuzbashyan2021} using Eq. \eqref{chempot} as well.
The time-dependence of the Loschmidt echo is plotted in Fig. \ref{fig:Uquench} (b). \jav{When the difference is large between the initial and final HK interaction strength, one can observe cusps, a non-analytic feature, in the lower period of the oscillations.}

%It is important to note that the time instants of the non-analyticity do not coincide with the critical times of the order parameter similarly to Ref. \cite{vajna2014}. \jav{Here, the critical times of the order parameter are meant as the time instants where the order parameter passes through zero.}

To study the non-analyticities systematically, we define from the numerical values $f_i = \ln\left(\mathcal{F}(t_i)\right)/N$ the discrete second derivative
\begin{gather}
s_i = \frac{f_{i-1} - 2f_i + f_{i+1}}{\Delta \tau^2}
\end{gather}
where $\Delta \tau$ is the time step in the numerical simulation.
As a continuous function, the Loschmidt echo would show non-analyticities \jav{where the first derivative exhibits a finite jump and, hence, the second derivative is not defined}. Numerically, we claim that the Loschmidt echo non-analytic if there exists 
$i$ such that both $|s_i| > 2(|s_{i-1}| + |s_{i+1}|)$ and $|s_i|>0.1\Delta \tau$ is fulfilled. In Fig. \ref{fig:Uline}, the \jav{region shaded with lightblue} indicates where non-analytic behavior was found, i.e., where DQPT emerges. For the specific example of $g/W=0.2$ and $U_f/W=0.5$, the transition between non-analytic and analytic regimes is found at $U_i/W\approx 0.36$ 
and $U_i/W\approx 0.64$. Interestingly, these critical values slightly differ from those of the DPT indicated by the order parameter.
In particular, the green curve of Fig. \ref{fig:Uquench} corresponding to $U_i/W=0.35$ highlights clearly what we discussed above: while the order parameter approaches a constant, non-zero value in the long time limit, 
hence no DPT occurs, the Loschmidt echo  displays non-analytic behaviour, indicating the
occurrence of DQPT.

\section{Momentum space entanglement}
\label{sec:ent}

\begin{figure}[t!]
\centering
\includegraphics[width=8cm]{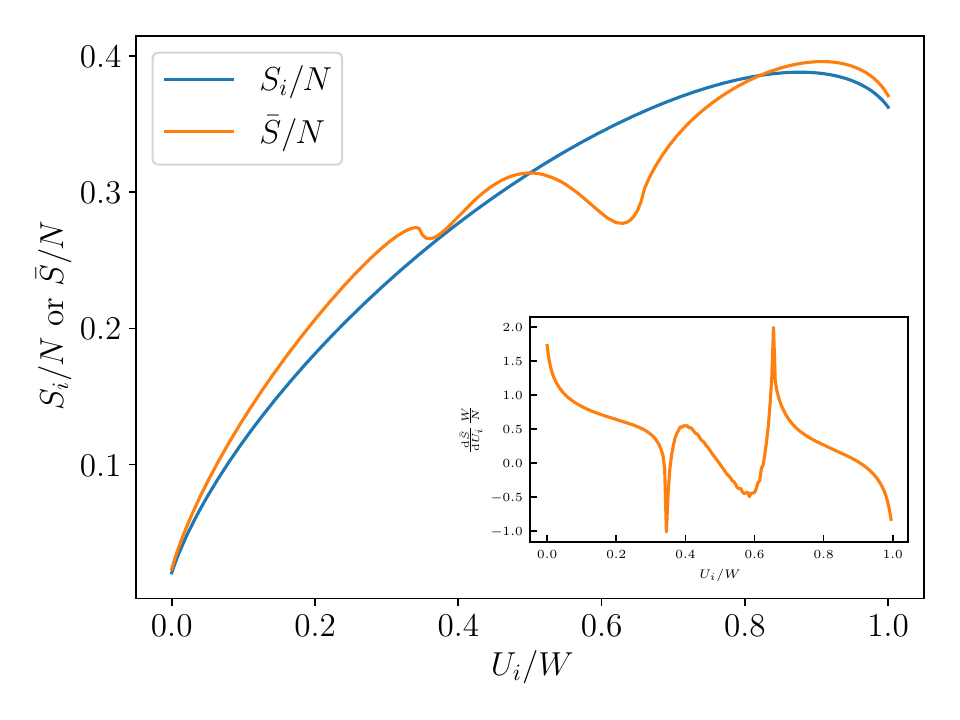}
\caption{The average momentum space entanglement entropy for $U_f/W=0.5$ between $k>0$ and $k<0$ modes is plotted for the initial state as well as in the steady state. The inset depict the interaction derivative of the entropy in the steady state, displaying sharp peaks
at the DPT points.}
\label{fig:UlineS}
\end{figure}

In order to better understand and characterize the properties of the steady state, we focus on its entanglement properties\cite{eisert,nielsen,amico}.
While the most common partitioning for entanglement is  spatial, i.e. done in real-space, other ways 
of partitioning are equally fruitful\cite{thomale}. 
Cooper pairs are made of particles with opposite momentum therefore characterizing their entanglement properties in momentum space is expected to contain all essential information about
the non-local quantum correlations and superconductivity. 
This is defined as
\begin{gather}
S(t) = -\Tr\left[\rho_+(t) \ln \rho_+(t)\right]
\end{gather}
where $\rho_+(t)=\Tr_-\left[|\Psi(t)\rangle \langle \Psi(t)|\right]$ is the reduced density matrix for $k>0$ after tracing out the $k<0$ modes. 
By using the coefficients of the wavefunction, the momentum space entanglement entropy is
\begin{gather}
S(t) =- \sum_{k>0} \left(|x_k(t)|^2 \ln|x_k(t)|^2 + |y_k(t)|^2 \ln \left(\frac{|y_k(t)|^2}{2}\right) +\right. \nonumber  \\
\left. + |z_k(t)|^2 \ln|z_k(t)|^2 \right)\,.
\label{eq:ent}
\end{gather}
The entanglement is computed based on the numerical solutions for $x_k$, $y_k$ and $z_k$ and is shown in Fig. \ref{fig:UlineS}. Here, $S_i$ corresponds to the initial state entanglement, while
$\bar{S}$ stands for the late time average of the  entanglement entropy.

The momentum space entanglement entropy satisfies a volume law even for the ground state.
We find that this entanglement does not change significantly between the initial and final state, though small enhancement or suppression is observed depending on the initial and final HK interaction.
The most notable feature is that this entanglement is large for large initial $U_i$ when the late time superconducting order parameter vanishes. Even though superconductivity is absent from
this steady state, the entanglement between modes with positive and negative momenta is very large for large initial HK interaction. 
The derivative of the entanglement with respect to the change of the HK interaction, i.e. $\frac{\mathrm{d}\bar{S}}{\mathrm{d}(U_i-U_f)}$ (which is  $\frac{\mathrm{d}\bar{S}}{\mathrm{d} U_i}$ for the current protocol) 
displays sharp peaks when DPT happens. That is, it signals
the late time vanishing of superconductivity.

\section{Conclusion}
In this paper, we studied the time evolution following a sudden quench of the HK interaction $U$ in the superconducting HK model by leaving the superconducting coupling intact.
The time evolution of the order parameter exhibits three distinct behaviours depending on the value of the initial and final value of the interaction, $U_i$ and $U_f$. 
We find that for small (intermediate) quenches, the superconducting gap exhibits undamped (damped) oscillation around a finite value while for larger quenches, the gap converges to zero in an oscillating fashion. The latter is regarded as one type of DPT.

In addition, we also inspected the time dependence of the Loschmidt echo, corresponding to the overlap of the initial and time evolved wavefunctions. 
Non-analytic temporal behaviour is identified for large quenches, though the "phase boundary" for this to happen
\jav{is close} to what is identified from the long time dynamics of the order parameter \jav{but the boundaries do not occur at the same critical value of $U$}. Therefore, DPTs and DQPTs do not coincide.

By investigating the momentum space entanglement between positive and negative energy modes, responsible for superconductivity, we find that this entanglement remains very large even for
regions with vanishing superconducting order parameter, corresponding to large initial value of the HK interaction. The derivative of the  momentum space entanglement with respect to HK interaction in the steady state clearly signals the vanishing 
amplitude of the superconducting gap by developing sharp peaks at the critical value of HK interaction, corresponding to DPT.

\begin{acknowledgments}
We acknowledge the support of the National Research, Development and Innovation Office - NKFIH  Project Nos. K134437 and 
K142179. \' A.B. acknowledges the support of the Slovenian Research and Innovation Agency (ARIS) under P1-0416 and J1-3008.
\end{acknowledgments}

\bibliographystyle{apsrev}
\bibliography{hk}

%\pagebreak 
\appendix
\section{HK quench with final interaction strength $U_f/W=0.3$}
\begin{figure}[h!]
\centering
\includegraphics[width=8cm]{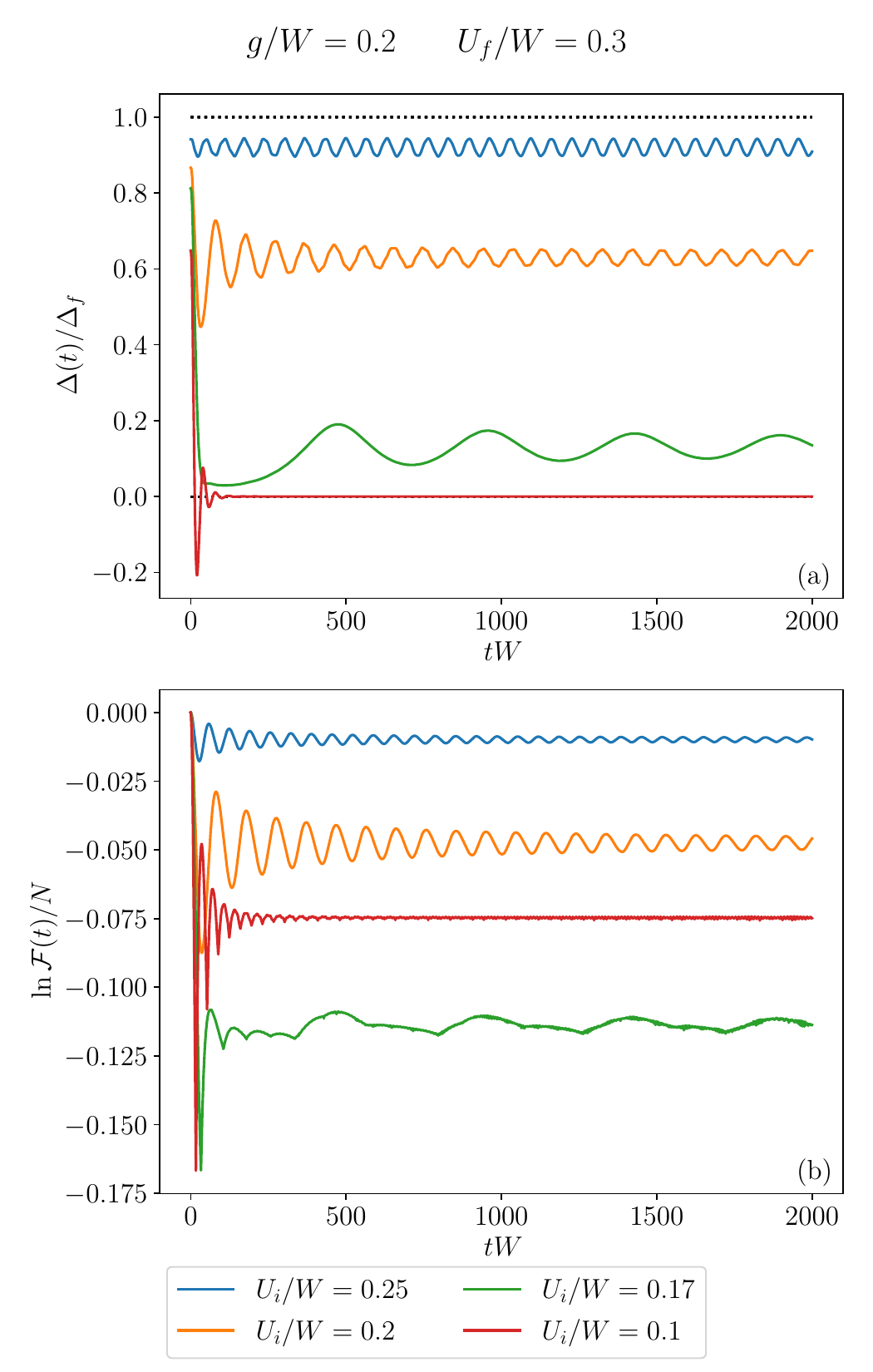}
\caption{Time evolution of the order parameter, $\Delta(t)$, and the Loschmidt echo, $\mathcal{F}(t)$ for various initial interaction strength, $U_i$ and for $U_f/W=0.3$.}
\label{fig:Uquench0.3}
\end{figure}

\begin{figure}[h!]
\centering
\includegraphics[width=8cm]{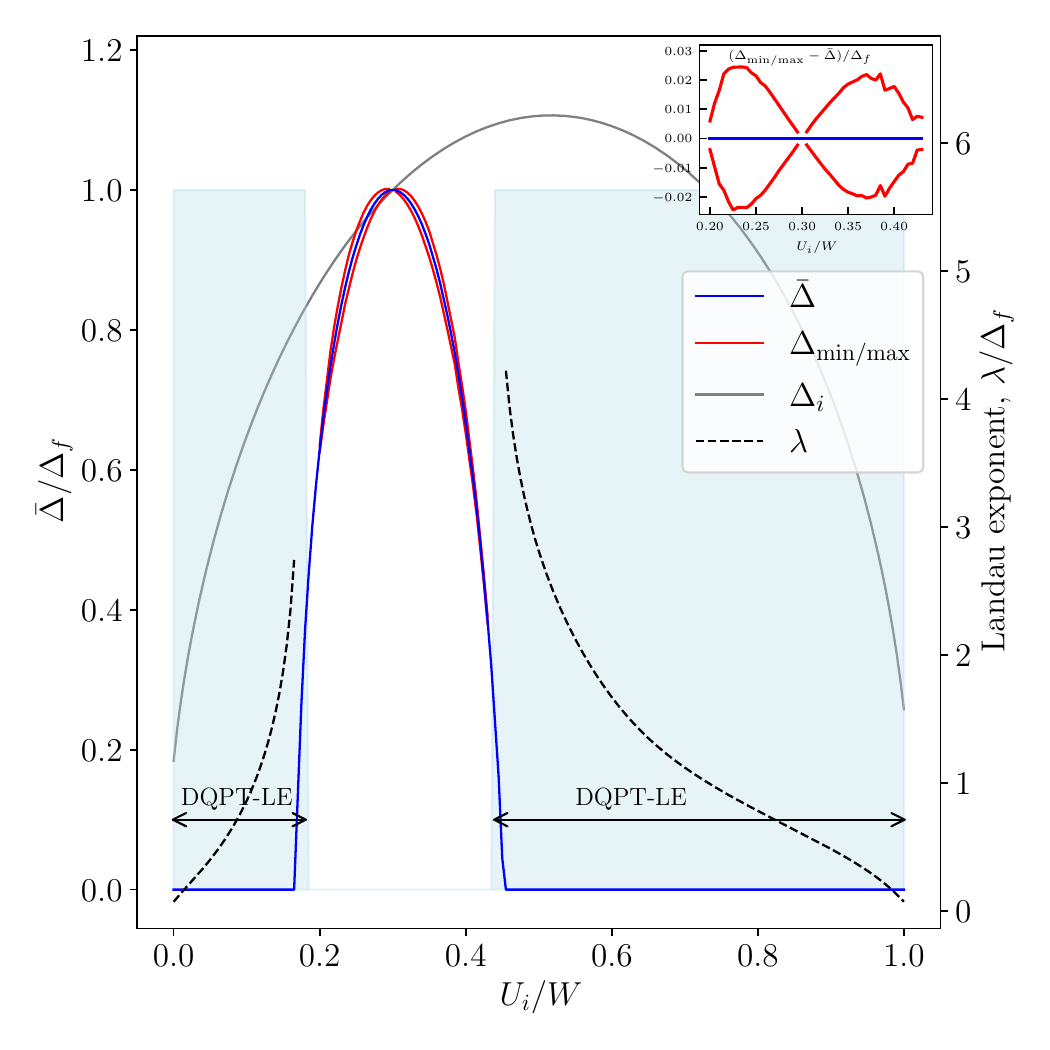}
\caption{The average order parameter (solid, blue line) for $U_f/W=0.3$ as a function of initial interaction strength $U_i$ measured on the left vertical axis, together with its maximal and minimal value in the case
of persistent oscillation. 
The right vertical shows Landau damping parameter (dashed, gray line) in the region when the order parameter vanishes in the long time limit. 
The hatching indicates the regime where the Loschmidt echo exhibits non-analytic features. 
For the plot, only the time instants after $t_\mathrm{cutoff} = 300 /W$ were taken into account.
The inset zooms into the region of persistent oscillation for small quenches.}
\label{fig:Uline0.3}
\end{figure}

\begin{figure}[h!]
\centering
\includegraphics[width=8cm]{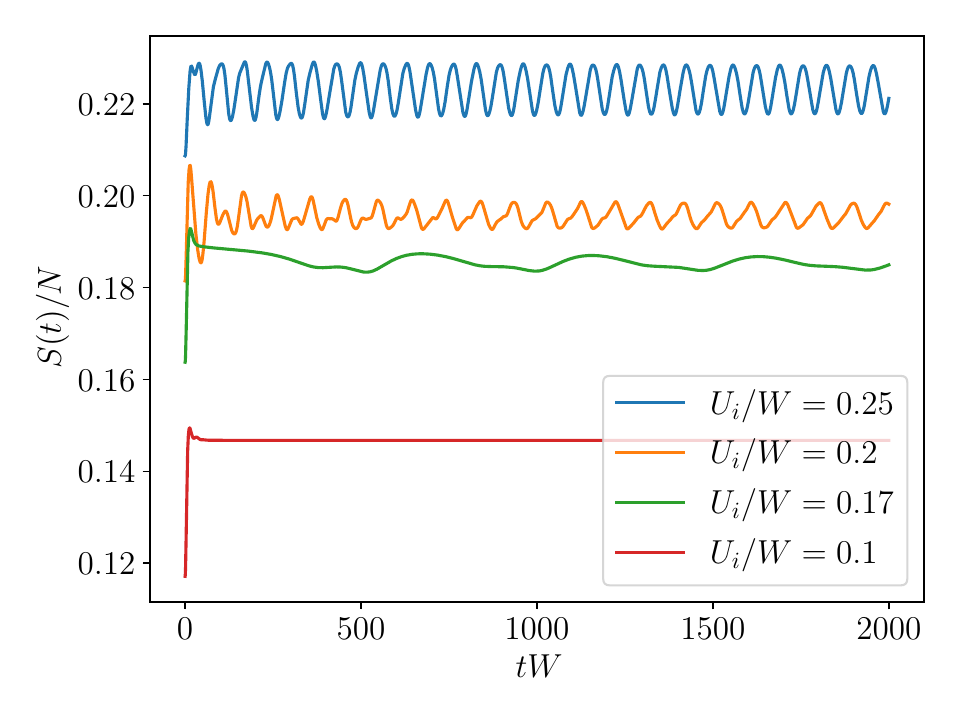}
\caption{Time evolution of the momentum space entanglement for various values of the initial HK interaction and for the final interaction strength $U_f/W=0.3$.}
\label{fig:ent}
\end{figure}

\begin{figure}[h]
\centering
\includegraphics[width=8cm]{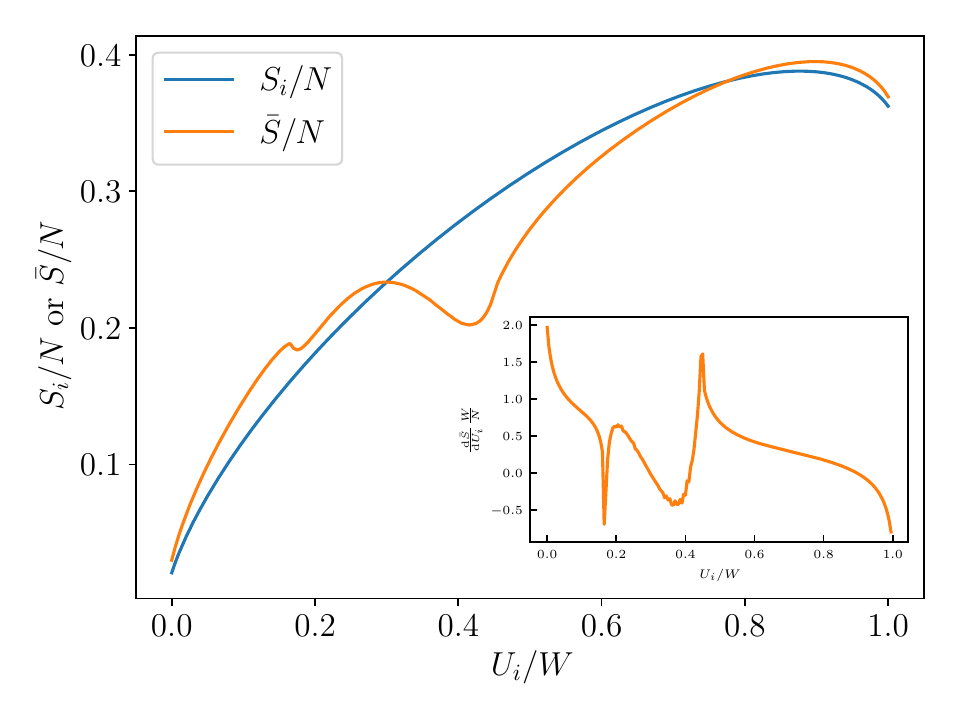}
\caption{The average momentum space entanglement entropy for $U_f/W=0.3$ between $k>0$ and $k<0$ modes is plotted for the initial state as well as in the steady state. The inset shows the interaction derivative of the steady state entropy.}
\label{fig:UlineS0.3}
\end{figure}

To demonstrate qualitatively similar dynamical features of the HK model with different final interaction strength, we perform numerical calculations for $U_f/W=0.3$. The typical time evolution of the order parameter is plotted in Fig. \ref{fig:Uquench0.3}
The results displays again the three distinct behaviors, namely the persistent oscillation, the damped oscillation and the vanishing behavior.
By sweeping the initial HK interaction between zero and $W$, we obtain the phase diagram as shown in Fig. \ref{fig:Uline0.3}, which  exhibits 
qualitatively similar features to  the case of $U_f/W=0.5$. The only difference is in the location of the different regions.
Finally, we compute the momentum space entanglement based on Eq. \eqref{eq:ent} of the main text. The time evolution of the entanglement is shown in Fig. \ref{fig:ent}. We also study the initial and the steady state entanglement 
as a function the initial HK interaction, see Fig. \ref{fig:UlineS0.3}, and find similar features as for $U_f/W=0.5$.


\end{document}